\documentstyle[12pt]{article}

\begin{document}
\baselineskip 22pt

\vspace{1.5cm}
\begin{center}
\begin{Large}
\begin{bf}
New Type of Vector Gauge Theory \\ from Noncommutative Geometry  \\
\vspace{15mm}
\end{bf}
\end{Large}

\vspace{2mm}

{\large Chang-Yeong  Lee${}^{*}$} \\

{\it Department of Physics, Sejong University \\ Seoul
    143-747, Korea }\\

\vspace{10mm}
{\large {\bf Abstract}}\\
\end{center}

\noindent
Using the formalism of noncommutative geometric gauge theory 
based on the superconnection concept, 
we construct a new type of vector gauge theory possessing 
a shift-like symmetry and 
the usual gauge symmetry.
The new shift-like symmetry is due to the matrix derivative
of the noncommutative geometric gauge theory, and this gives rise to
a mass term for the vector field without introducing the Higgs field.
This construction becomes possible by using 
a constant one form even matrix for the matrix derivative, for which
 only constant zero form odd matrices have been used so far.
The fermionic action in this formalism is also constructed 
and discussed.
\\

\thispagestyle{empty}
\vfill


\noindent
\hbox to 10cm{\hrulefill}\\
\baselineskip 12pt
\indent
${}^{*}$
E-mail: leecy@phy.sejong.ac.kr
\\

\pagebreak

\baselineskip 22pt

\noindent
{\large {\bf I. Introduction}}\\

In 1990, Connes and Lott \cite{conlot}
showed that the standard model can be obtained from noncommutative 
geometric gauge theory. 
In this noncommutative framework, the Dirac K-cycle 
plays an important role and the Higgs mechanism is implemented by
a generalized Dirac operator acting on a discrete space.
In Ref. \cite{lee}, we showed that the same mechanism can be implemented
 via matrix derivative approach 
based on the superconnection concept \cite{qui,ns}.
There already have been many works along this line \cite{couqet,sch,lhn},
but in all these works only constant zero form odd matrices for the 
matrix derivative have been used in order to conform it 
to the Connes-Lott's generalized Dirac operator which
is in essence a matrix commutator with a constant odd matrix. 
In Ref. \cite{lee}, we found the conditions that the matrix derivative
should satisfy when the noncommutative geometric gauge theory is constructed
via the superconnection formalism.
There we found another type of the matrix derivative consisting 
of constant(closed) one form matrix which has nonvanishing elements 
in the even part and vanishing elements in the odd part.
In this paper, we use this one form even matrix for the matrix derivative,
and construct a new type of vector gauge theory possessing  
 the usual gauge symmetry and  a shift-like
symmetry which in turn gives rise to
a mass term for the vector gauge field without 
recourse to the Higgs field. 
\\

\noindent
{\large {\bf II. Noncommutative geometric gauge theory in the
 superconnection formalism  }}\\

Superconnection was first introduced in mathematics by Quillen in
1985 \cite{qui}. However, in physics this concept was used earlier by
Thierry-Mieg and Ne'eman without giving it a name in 1982 \cite{tmn} 
under the notion of a generalized connection {\it a la} Cartan \cite{crt}. 
Then in 1990, Ne'eman and Sternberg \cite{ns} applied superconnections 
for the Higgs mechanism.

Let $V = V^{+} \oplus V^{-}$ be a super (or $Z_{2}$-graded) complex vector
space, then the algebra of endomorphisms of $V$ is a superalgebra with
the even or odd endomorphisms.
Let ${\cal E}={\cal E}^{+} \oplus {\cal E}^{-}$ be
a super (or $Z_{2}$-graded) vector bundle over a manifold $M$, and
$\Omega(M) =\oplus \Omega^k(M)$ be the algebra of smooth differential
forms with complex coefficients.
Then, the space of ${\cal E}$ valued
differential foms on $M$, $\Omega(M, {\cal E})$,
 has a $Z \times Z_{2}$ grading, and here we are
mainly concerned with its total $Z_{2}$ grading. 

In Ref. \cite{lee}, we showed that a generalization of the superconnection
concept can yield the matrix derivative of the noncommutative
geometric gauge theory \cite{sch}.
There, the generalized superconnection 
is given by \cite{lee}
\begin{equation}
 \bigtriangledown =  {\bf d}_{t} + \omega , 
\end{equation}
where
 $ \; {\bf d}_{t}= {\bf d}+{\bf d}_{M}$ is a generalization of the
one form exterior derivative satisfying the derivation property, 
and  $\omega$ is a generalized connection given by
$\omega = \left( \begin{array}{cc} \omega_{0} & L_{01} \\
  L_{10}  & \omega_{1}  \end{array} \right).$ Here,
$\omega_{0}, \ \omega_{1}$ are matrices of odd degree differential forms
and  $ L_{01}, \  L_{10}$ are matrices of even degree differential forms.
The multiplication rule is given by
\begin{equation}
 (u \otimes a)\cdot (v \otimes b) = (-1)^{\vert a \vert  \vert v \vert}
 (uv)\otimes (ab), \; \; \; u,v \in \Omega(M), \; \; a,b \in {\cal A},
\end{equation}
where ${\cal A}$ is the endomorphisms of $V$.

In the matrix representation,  ${\bf d} =\left( 
\begin{array}{cc} d & 0 \\ 0  & d  \end{array} \right) $ 
where  $d$ inside
the matrix denotes the usual 1-form exterior derivative 
times a unit matrix, and 
${\bf d}_{M}$ is given below.
Since ${\bf d}_{M}$ should behave as a part of the superconnection operator
\cite{bgv} in a sense, we write it as a (graded) commutator operator
\begin{equation}
{\bf d}_{M}= \left[ \eta, \; \cdot \ \right], \; \; \; \eta \in
   \Omega (M, {\cal E}).
\end{equation}
Now,  ${\bf d}_{M}$ should satisfy  
\begin{eqnarray}
  \; \; {\bf d}_{M}^2 &=& 0, \; \; \; \; \; \; \;
   {\bf d}{\bf d}_{M} + {\bf d}_{M}{\bf d}=0, \label{c11} \\
 {\bf d}_{M}(\alpha \beta)& =& ({\bf d}_{M} \alpha) \beta +
(-1)^{\vert \alpha \vert} \alpha ({\bf d}_{M} \beta), \; \; \; \
 \alpha, \beta \in \Omega (M, {\cal E}). \nonumber
\end{eqnarray}
In Ref. \cite{lee}, two simple solutions satisfying
the  above conditions were given by 
\\
 (1)  $ ~ \;  \eta =   \left( \begin{array}{cc} u & 0 \\
 0  & v  \end{array} \right) $ where $u, \; v$ are odd degree closed forms
with their coefficient matrices satisfying $ u^2 = v^2 \propto 1$ or 
$ u^2=v^2=0,$ \\
 (2) $ \; \eta =   \left( \begin{array}{cc} 0 & m \\
 n  & 0  \end{array} \right)$ where $m, \; n$ are even degree closed forms
with their coefficient matrices satisfying $mn=nm \propto 1$ or 
 $mn=nm=0$. \\
If we take the second solution with 0-form $m, \; n$, then this
choice yields the so-called matrix derivative \cite{sch}
 ${\bf d}_{M}= [\eta , \ \cdot \ ] $
with $ \eta =
 \left( \begin{array}{cc} 0 & \zeta \\ \overline{\zeta}  & 0
\end{array} \right)$ where $  \zeta,  \; \overline{\zeta}$ are
 0-form constant matrices satisfying $  \zeta \overline{\zeta} =
\overline{\zeta} \zeta \propto 1 $.

With the use of the generalized superconnection, the
 curvature is now given by
\begin{equation}
 {\cal F}_{t}=({\bf d}_{t} + \omega )^2 = {\bf d}_{t} \omega 
   + \omega^2 .  
\end{equation}
In this formulation, the Yang-Mills action is given by  
\begin{equation}
 I_{YM} =  \int_{M} {\rm Tr} 
  ( {\cal F}_{t}^{\star} \cdot {\cal F}_{t} ) 
\label{c8}
\end{equation}
where $\star$ denotes taking dual for each entries of ${\cal F}_{t}$ 
as well as taking Hermitian conjugate.  
The fermionic action is given by
\begin{equation}
I_{sp} =\int_{M} \overline{\Psi} \gamma^{\mu}
 ({\bf d}_{t}  + \omega  )_{\mu} \Psi, \; \; \;
\Psi \in V \otimes S
\label{c9}
\end{equation}
where $S$ is a spinor bundle.
\\

\noindent
{\large  \bf III. Massive vector gauge theory with unbroken symmetry  }\\

Now, we consider the first solution for ${\bf d}_{M}$ given 
in the previous section with 
$ \eta =  \left( \begin{array}{cc} \sigma  & 0 \\
 0  & \sigma'  \end{array} \right) $
where $\sigma, ~ \sigma'$ are constant 1-form matrices whose 
squares are either proportional to a unit matrix or zero.
For the generalized connection $\omega$, we set
\begin{equation}
 \omega =  \left( \begin{array}{cc} A & 0 \\
 0  & A'  \end{array} \right)
\end{equation}
where $A, \ A' $ consist of one forms only.

For a definite understanding, we consider the case where $A, \ A'$ are
SU(2) valued 1-form fields, and $\sigma$ and $\sigma'$ are
proportional to a SU(2) Pauli matrix, say $ \tau_3$:
\begin{eqnarray}
 & & A = \frac{i}{2} A^a_{\mu} \tau_a dx^{\mu} \equiv A_{\mu} dx^{\mu}, 
~~~~ A'= \frac{i}{2} {A'}^a_{\mu} \tau_a dx^{\mu} \equiv {A'}_{\mu} dx^{\mu},
  \nonumber   \\ 
 & & 
\sigma = \sigma' =  \frac{i}{2} m  \tau_3 n_{\mu} dx^{\mu}
 \equiv \sigma_{\mu} dx^{\mu}. \label{nc9}
\end{eqnarray}
Here $\tau$'s are Pauli matrices, $n_{\mu}$ is a constant
four vector, and
$m$ is a constant parameter.
Throughout the paper, we use the metric $g_{\mu \nu}=(-1,+1,+1,+1)$ and 
$\epsilon_{0123} = +1$, and the wedge product between forms is understood.

The curvature  
\begin{eqnarray}
 {\cal F}_{t} & = & {\bf d}_{t} \omega + \omega^2  \\
      & = & {\bf d}\omega + [ \eta , \omega ]_{\pm} +  \omega^2
 \nonumber
\end{eqnarray}
is now given by
\begin{eqnarray*}
 {\cal F}_{t} & = & \left(
\begin{array}{cc} d & 0 \\ 0  & d  \end{array} \right)
\left( \begin{array}{cc} A & 0 \\
 0  & A'  \end{array} \right) + \left[ 
 \left( \begin{array}{cc} \sigma  & 0 \\
 0  & \sigma'  \end{array} \right) , 
 \left( \begin{array}{cc} A & 0 \\
 0  & A'  \end{array} \right) \right]_{+} + 
 \left( \begin{array}{cc} A & 0 \\
 0  & A'  \end{array} \right) \cdot \left( \begin{array}{cc} A & 0 \\
 0  & A'  \end{array} \right).
\end{eqnarray*}
The first and third terms are the usual ones and the second term is a new
piece due to the matrix derivative which we calculate below.
Since all the odd parts are vanishing, the upper and lower diagonal 
parts do not mix. Hence, we mostly consider the upper part in our 
calculation. 
\\
Now, 
\begin{eqnarray} 
 \sigma  A + A \sigma 
 &  = &  \frac{1}{2} \left\{ [\sigma_{\mu}, A_{\nu}] 
   - [ \sigma_{\nu}, A_{\mu} ] \right\} dx^{\mu}dx^{\nu}  \nonumber \\
 & = & - \frac{1}{4} m \ n_{ [ \mu} \left( \begin{array}{cc} 0 & A_1 -i A_2 \\
 -A_1 - i A_2  & 0  \end{array} \right)_{\nu ] }  dx^{\mu}dx^{\nu} 
 \nonumber \\
 & \equiv & \frac{1}{2} {\cal A}_{\mu \nu} dx^{\mu}dx^{\nu}. 
\end{eqnarray}
Thus the curvature is given by
\begin{equation}
{\cal F}_{t} = \left( \begin{array}{cc} F_t & 0 \\ 0  & F'_t
  \end{array} \right)
\end{equation}
with
\begin{equation}
 F_t =  \frac{1}{2} (F_{\mu \nu} + {\cal A}_{\mu \nu}) dx^{\mu}dx^{\nu},
\end{equation}  
and $ F'_t$ is the same as $F_t$ except that $A$ is replaced by $A'$,
and $ F_{\mu \nu}$ is the usual one,
\begin{equation}
F_{\mu \nu} =  \partial_{ [ \mu } A_{\nu ] } 
           + [  A_{\mu}  ,  A_{\nu} ]. 
\end{equation}

  Following the same calculational step as in Ref. \cite{lhn}, we 
obtain the Yang-Mills type action of massive gauge fields  
 from Eq. (\ref{c8}), 
\begin{eqnarray}
 I_{YM}& =& 
   \int_{M}{\rm Tr} ( {\cal F}_{t}^{\star} \cdot {\cal F}_{t} ) 
  \label{c16} \\
        & = &   \frac{1}{2} \int_{M} d^4 x \ {\rm Tr} \left[ \left( 
   F_{\mu \nu} + {\cal A}_{\mu \nu} \right) 
   \left( F^{\mu \nu} + {\cal A}^{\mu \nu} \right) +
      \left( {\rm terms ~ with} ~~ A \rightarrow A' \right) \right]
      \nonumber \\
  & = &   \frac{1}{2} \int_{M} d^4 x \ 
        {\rm Tr} \left[  F_{\mu \nu}F^{\mu \nu} +
   F_{\mu \nu}{\cal A}^{\mu \nu} + {\cal A}_{\mu \nu}F^{\mu \nu}
     + {\cal A}_{\mu \nu}{\cal A}^{\mu \nu}
    + \left( {\rm terms ~ with} ~~   A \rightarrow A' \right) \right].
 \nonumber
\end{eqnarray}
The fourth term provides quadratic terms homogeneous in $A_1$
 and $A_2$:
\begin{equation}
   \frac{1}{2} {\rm Tr} {\cal A}_{\mu \nu}{\cal A}^{\mu \nu}  = 
   \frac{1}{2} m^2 \left[ - n_{\mu}n^{\mu} \left( A_{1 \nu} A_{1}^{\nu} 
    + A_{2 \nu} A_{2}^{\nu} \right) +  n_{\mu}n_{\nu} 
           \left( A_{1}^{\mu} A_{1}^{\nu}
    + A_{2}^{\mu} A_{2}^{\nu} \right) \right] . 
\label{c17}
\end{equation}
The second and third terms also give terms quadratic in $A$ but mixed
in  $A_1$ and $A_2$:
\begin{equation}
 \frac{1}{2}{\rm Tr} \left(
 F_{\mu \nu}{\cal A}^{\mu \nu} + {\cal A}_{\mu \nu}F^{\mu \nu}
 \right) =  m \epsilon^{a b} \left( n_{\mu}A_{a \nu} 
               \partial^{\mu} A_b^{\nu} -
            n_{\mu}A_{a \nu}  \partial^{\nu} A_{b}^{\mu} 
         \right) + ~ O(A^3) 
\label{c18} 
\end{equation}
where $a,b=1,2$ and $\epsilon^{1 2} = - \epsilon^{2 1} = 1, 
~ \epsilon^{1 1}= \epsilon^{2 2} =0$.

Before we perform diagonalization and obtain the propagators for these fields,
we first identify the symmetry of the action.
In Ref. \cite{lhn}, the so-called horizontality condition was used
to analyze the BRST symmetry of the noncommutative geometric gauge theory.
Since we use the same superconnection framework, 
the BRST analysis will be more convenient 
for finding the symmetry of
the theory.
In the Yang-Mills theory, the horizontality condtion is given by
\cite{tm,bt,nt,ln}
\begin{equation}
\widetilde{F} \equiv \widetilde{d}\ \widetilde{A}\ + \widetilde{A}\
\widetilde{A}\ =F ,
\label{2c4}
\end{equation}
where
\begin{eqnarray*}
\widetilde{A}\ & =& A_{\mu} dx^{\mu} + A_{N}dy^{N}
+A_{\bar{N}}d {\bar{y}}^{\bar{N}}
 \equiv A + c  + \bar{c}  , \\
\widetilde{d} & =& d + s + \bar{s} , \; d = dx^{\mu}\partial_{\mu}, \;
s = dy^{N}
\partial_{N}, \; \bar{s} =d{\bar{y}}^{\bar{N}}\partial_{\bar{N}}, \\
F & = & dA+AA={1\over 2}F_{\mu\nu}dx^{\mu}dx^{\nu} .
\end{eqnarray*}
Here $ y$ and $\bar{y}\ $ denote the coordinates in the direction of
gauge orbit of the principal fiber whose structure-group is
${\cal G} \otimes {\cal G} $, and $c, \ \bar{c}$ are ghost and antighost
fields.
The above horizontality condition now yields the BRST and anti-BRST
transformation rules for the Yang-Mills theory.
\begin{eqnarray}
 (dx)^1(dy)^{1} & : &
sA_{\mu}=D_{\mu}c
\nonumber , \\
(dx)^1(d{\bar{y}})^{1} & : &
{\bar{s}}A_{\mu}=D_{\mu}{\bar{c}}
\nonumber , \\
(dy)^{2} & : &
s c =-c c
\label{2c5} , \\
(d{\bar{y}})^{2} & : &
{\bar{s}}{\bar{c}}=-{\bar{c}}{\bar{c}}
\nonumber , \\
(dy)^1(d{\bar{y}})^{1} & : &
s{\bar{c}}+{\bar{s}}c =-[ c ,{\bar{c}} ] .
\nonumber
\end{eqnarray}
In the superconnection framework, the horizontality condition is given
as follows \cite{lhn}.
\begin{equation}
  \widetilde{{\cal F}_t} \equiv \widetilde{{\bf d}_t}
\widetilde{\omega} +\widetilde{\omega}  \cdot
\widetilde{\omega} = {\cal F}_t  
\label{c21}
\end{equation}
where
\begin{eqnarray}
 \widetilde{{\bf d}_t} & = &  {\bf d}_{t} + {\bf s} + \bar{\bf s} ,
 \label{c22} \\
\widetilde{\omega} & = & {\omega} + {\cal C} + \bar{\cal C},
\label{c23} 
\end{eqnarray}
and 
\begin{equation}
{\bf s}  =   \left(\matrix{s&0\cr 0&s\cr}\right), ~~
\bar{\bf s}  =  \left( \begin{array}{cc} \bar{s}  & 0 \\
    0 & \bar{s} \end{array} \right), ~~
{\cal C} = \left(
         \begin{array}{cc} c & 0 \\ 0 & c' \end{array} \right),
~~ \bar{\cal C} = \left(
        \begin{array}{cc} {\bar{c}}  & 0 \\ 0 & {\bar{c'}} \end{array}
        \right).
\label{c24}
\end{equation}
The above horizontality condition yields the following BRST and anti-BRST
transformation rules:  
\begin{eqnarray}
(dy)^{1} & : &
{\bf s} {\omega}  =  -{\bf d}_{t} {\cal C}
- {\omega}  \cdot {\cal C} - {\cal C} \cdot  {\omega} , \label{c25}  \\
(d{\bar{y}})^{1} & : &
\bar{\bf s}  {\omega}  =  -{\bf d}_{t}  \bar{{\cal C}}
-  {\omega} \cdot \bar{{\cal C}} -\bar{{\cal C}} \cdot  {\omega},  \\
(dy)^{2} & : &
{\bf s}  {\cal C}  =  -{\cal C} \cdot {\cal C} ,  \label{c26} \\
(d{\bar{y}})^{2} & : & \bar{\bf s} \bar{{\cal C}}   =
-\bar{{\cal C}} \cdot \bar{{\cal C}} ,  \\
(dy)^1(d{\bar{y}})^{1} & : &
{\bf s} \bar{{\cal C}}  + \bar{\bf s} {\cal C}  +{\cal C}
\cdot \bar{{\cal C}}
+ \bar{{\cal C}} \cdot {\cal C}   =  0 .
\end{eqnarray}
Since all the odd parts vanish as before, we again consider only the upper
diagonal (even) parts in our calculation. Then, the BRST
and anti-BRST transformation rules for the fields appearing in the
upper parts can be written as    
\begin{eqnarray}
 & &   s A  =  -d c - [ \sigma , c ]_{+} - [A, c]_{+} , \label{c30} \\ 
 &  & \bar{s} A  =  - d  \bar{c}  - [ \sigma ,\bar{c} ]_{+} 
      -[A, \bar{c} ]_{+} ,  \\
 & &  s  c  =   -c c ,  \label{c32} \\
 & &  \bar{s} \bar{c}   =  -\bar{c} \bar{c} ,  \\
 &  &  s \bar{c}  + \bar{s} c  + c \bar{c} + \bar{c} c   =  0 , 
\end{eqnarray}
where
\[ c=  \frac{i}{2} c_a \tau^{a}, ~~ 
   \bar{c} = \frac{i}{2} \bar{c}_a \tau^{a}, ~~ a=1,2,3 . \] 
Now, one can check that the above BRST and anti-BRST transformations
are nilpotent, $ s^2 = \bar{s}^2 = 0$, and the total curvature 
$F_t=d A + A A + \sigma A + A \sigma $ transforms as the usual
curvature $F=dA + AA$, 
\begin{equation}
s F_t = -[c, F_t].
\label{c35}
\end{equation}
Therefore, our Yang-Mills action, 
\begin{equation}
 I_{YM}^0  =  \int_{M} {\rm Tr} F_t^* F_t  
          =  \frac{1}{2} \int_{M} d^4 x \ {\rm Tr} \left[ \left(
   F_{\mu \nu} + {\cal A}_{\mu \nu} \right)
   \left( F^{\mu \nu} + {\cal A}^{\mu \nu} \right) \right] 
\label{act}
\end{equation}
where $*$ denotes the Hodge dual,
is invariant under the above given BRST(anti-BRST) transformation.
Since the BRST and gauge transformations
for classical fields are the same except for a switch between the
 classical gauge parameter
and the ghost field, one can check that the action (\ref{act})
 is invariant under the following
gauge transformation
\begin{equation}
\delta A_{\mu} = \partial_{\mu} \varepsilon + [ A_{\mu}, \varepsilon] 
    + [\sigma_{\mu}, \varepsilon]
\label{c36}
\end{equation}
where
$\varepsilon = \frac{i}{2} \varepsilon_a \tau^a ~~ (a=1,2,3)$ is a zero 
form gauge parameter.  
In order to obtain the propagators we use the following gauge fixing 
term for the action (\ref{c16})
\begin{equation}
{\cal L}_{g.f.} = \frac{1}{\xi} {\rm Tr} ( {\bf d}_{t} \omega^{\star})^2,
\end{equation}
which is translated into the following condition for the action (\ref{act}), 
\begin{equation}
{\cal L}_{g.f.}^{0} = \frac{1}{\xi} {\rm Tr} (\partial_{\mu} A^{\mu} +
    \sigma_{\mu} A^{\mu} - A_{\mu} \sigma^{\mu} )^2
\label{gfe}
\end{equation}
where $\sigma_{\mu}, ~A_{\mu}$ are given by Eq. (\ref{nc9}).
In terms of 
\begin{equation}
W_{\pm}^{\mu}= \frac{1}{\sqrt{2}}(A_1^{\mu} \mp i A_2^{\mu}),
\label{wdf}
\end{equation}
we obtain the propagators for $W_{\pm}, ~ A_3$, after some calculation:
\begin{eqnarray}
W_{\pm} & : & \bigtriangleup^{\pm}_{\mu, \nu} =
  \frac{1}{(P^{\pm})^2}\left( g_{\mu \nu} + (\xi -1) \frac{P^{\pm}_{\mu}
           P^{\pm}_{\nu}}{(P^{\pm})^2} \right),
\nonumber \\
A_3 & : & \bigtriangleup^{3}_{\mu, \nu} =
   \frac{1}{p^2} \left( g_{\mu \nu} + (\xi -1) \frac{p_{\mu} p_{\nu}}
      {p^2}  
 \right),  \label{prop}
\end{eqnarray}
where $P^{\pm}_{\mu} = p_{\mu} \pm m n_{\mu}$.
If we set $n \cdot p = n_{\mu} p^{\mu} = 0$, then the denominator of the
$W$-propagator becomes $(P^{\pm})^2 = p^2 + m^2 n^2$. 
Thus, choosing $n_{\mu}$ satisfying the above condition exhibits
the $W$ field having a mass; $({\rm mass})^2=m^2 n^2$.
\\

\noindent
{\large \bf IV. Fermionic action }\\

 The femionic action Eq. (\ref{c9}) shows that the spinor 
 $ \Psi$ belongs to a $Z_2$-graded vector space, thus we write
 $\Psi = (\psi_{+}, \psi_{-})$. Then the fermionic action, 
\begin{equation} 
I_{sp}  =   \int_{M} d^4 x \  \left[ \overline{\Psi}
     \gamma^{\mu} \left( 
  {\bf d}_{\mu}  +   \omega_{\mu} \right) \Psi + \overline{\Psi}
      \gamma^{\mu} \eta_{\mu} \Psi \right] 
\end{equation}
where $ {\bf d}_{\mu}= \left( \begin{array}{cc} \partial_{\mu}  & 0 \\
 0  & \partial_{\mu}  \end{array} \right)$ and
 $\eta_{\mu}= \left( \begin{array}{cc} \sigma_{\mu}  & 0 \\
 0  & {\sigma}_{\mu}  \end{array} \right)$,
can be written as 
\begin{equation}
I_{sp}  =   \int_{M} d^4 x \  \left[ \overline{\psi}_{+}
 \gamma^{\mu} \left( \partial_{\mu} + A_{\mu} \right) \psi_{+}
  +  \overline{\psi}_{+}  \gamma^{\mu} \sigma_{\mu} \psi_{+}
   + \left( {\rm terms ~ with~~} 
 + \rightarrow - \right) \right].
\end{equation}
Here, we assume $A=A'$ for convenience, and 
$\sigma_{\mu}= \frac{i}{2} m \tau_{3} n_{\mu}$ as before.
The action given above is a massless fermionic action except for the term,
$\overline{\psi}_{+}  \gamma^{\mu} \sigma_{\mu} \psi_{+}$, which looks 
similar to the usual mass term.
In order to compare it with the usual mass term,
$m \left( \overline{\psi}_L \psi_R + \overline{\psi}_{R} \psi_L \right)$,
we now set $\psi_+ = \psi_L= \frac{1+ \gamma_5}{2} \psi$ and 
$\psi_- = \psi_R = \frac{1-\gamma_5}{2} \psi$.
Then 
\begin{eqnarray}
  \overline{\psi}_{+}  \gamma^{\mu} \sigma_{\mu} \psi_{+} & = &
  \psi^\dagger (\frac{1+ \gamma_5}{2})^\dagger \gamma_0 \gamma^\mu
 \sigma_{\mu} (\frac{1+ \gamma_5}{2}) \psi
= \psi^\dagger \gamma_0  \gamma^\mu
 \sigma_{\mu} (\frac{1+ \gamma_5}{2})(\frac{1+ \gamma_5}{2})\psi \nonumber
 \\ &  = &
 \overline{\psi}  \gamma^\mu
 \sigma_{\mu} (\frac{1+ \gamma_5}{2}) \psi,
\label{fm1}
\end{eqnarray}
and the same way
\begin{equation}
\overline{\psi}_{-} \gamma^{\mu} \sigma_{\mu} \psi_{-} =
 \overline{\psi}  \gamma^\mu
 \sigma_{\mu} (\frac{1 - \gamma_5}{2}) \psi.
\label{fm2}
\end{equation}
Thus the two terms containing $\sigma_{\mu}$ 
become 
\begin{equation}
\overline{\psi}_{+}  \gamma^{\mu} \sigma_{\mu} \psi_{+}
 + \overline{\psi}_{-} \gamma^{\mu} \sigma_{\mu} \psi_{-}
= \overline{\psi}  \gamma^\mu
 \sigma_{\mu} \psi.
\end{equation}
Notice that $\sigma_{\mu}= \frac{i}{2} m \ n_{\mu} \tau_3$ commutes
with $\gamma$'s and only acts on the fundamental representaion of $\psi$,
say a doublet $ (  u, ~ d ) $, and $\gamma$'s act on $u$ or $d$ itself.
Thus this term has an extra factor of $\gamma^\mu  n_{\mu}$ compared
with the usual mass term.
\\

\noindent
{\large \bf V. Concluding remarks}\\

In this paper, we constructed a new type of vector gauge theory 
which possesses both the usual gauge symmetry and a shift-like
symmetry which gives rise to a mass term for the vector
gauge field.
So far, in the noncommutative geometric gauge theory,
only zero form constant odd matrices have been used for the 
matrix derivative, and this constant zero form odd matrix 
together with scalar fields appearing in the odd
part of the gauge multiplet (or superconnection) give rise to 
the Higgs mechanism. In the Connes-Lott formalism, this constant
odd matrix does the role of the generalized Dirac operator acting
on a discrete space.
However, constructing the noncommutative geometric gauge theory 
in the superconnection formalism, it is possible to use 
a constant one form even matrix for the matrix derivative \cite{lee}.
This in turn makes it possible to 
construct a massive vector gauge model similar to the Proca's while
maintaining gauge symmetry. 
It is also possible to provide mass terms for all the
gauge fields by properly choosing this constant one form even matrix.
For instance, if we replace $\tau_3$ appearing 
in $\sigma_{\mu}= \frac{i}{2} m n_{\mu} \tau_3$ with $\tau_1 +
\tau_2 + \tau_3$, then all $A_1, A_2, A_3$ fields become massive.
However, if we use an identity matrix for  $\sigma_{\mu}$, then 
there will be no massive vector field.

Finally, we would like to mention two things which looked confusing
in the previous sections.
In Eq. (\ref{prop}), the propagators for $W_{\pm}$ look
apparently different by a term involving the gauge fixing parameter $\xi$.
This difference can be removed for the choice of $\xi=1$, and for other
$\xi$ values we expect that
the contribution from this $\xi$ related term be cancelled by 
that of ghosts since the gauge fixing should not affect the
physics.\\
In the case of the fermionic action, the presence of the extra factor
$ \gamma^{\mu} n_{\mu}$ in the 
``mass" term looks puzzling.
However, the propagator for the $\psi$ field can be expressed
as 
\[ \bigtriangleup= \frac{1}{-ip_{\mu} \gamma^{\mu} + \sigma_{\mu}
      \gamma^{\mu}} = \frac{ ip_{\mu} \gamma^{\mu} + \sigma_{\mu}
      \gamma^{\mu}} { p^2 + \sigma^2},
\]
and thus exhibits that $\psi$ has a mass; $({\rm mass})^2 = \sigma^2$.

Now, we leave the more rigorous work on these issues along with 
the investigation of the
quantum effect for a future work.
\\

\noindent
{\large \bf Acknowledgments}\\
I would like to thank D.S. Hwang for helpful discussions,
B. Lee, W.T. Kim, P. Orland for helpful comments, and Y. Kiem for
reading the manuscript and comments.
This work was supported
in part by the Ministry of
Education, BSRI-97-2442, and KOSEF grant 971-0201-007-2.
\\

\end{document}